\documentclass[
    nobibnotes, aps, preprint, showkeys,
    superscriptaddress,
    floatfix, obeyspaces, spaces,
]{revtex4-2}

\usepackage{lmodern}                
\usepackage[T1]{fontenc}            
\usepackage[utf8]{inputenc}         

\usepackage{graphicx}               
\usepackage{booktabs}               
\usepackage{setspace}               
\usepackage[
    colorlinks,
    citecolor=blue,
    urlcolor=blue,
    filecolor=blue,
    linkcolor=black
]{hyperref}

\makeatletter
\def\figurename{Figure}
\def\tablename{Table}
\renewcommand*{\fnum@figure}{{\normalfont\bfseries\figurename~\thefigure}}
\renewcommand*{\fnum@table}{{\normalfont\bfseries\tablename~\thetable}}
\def\@caption@fignum@sep{\space}
\AtBeginDocument{%
  \def\@hangfrom@section#1#2#3{\@hangfrom{#1#2}{#3}}%
  \def\@hangfroms@section#1#2{#1{#2}}%
  \let\place@bibnumber\place@bibnumber@sup%
  \let\@cite\NAT@citesuper%
}
\makeatother

\usepackage{soul}                   
\usepackage{color}                  
\usepackage{xspace}                 

\begin{document}

\title[]{
    A roadmap for polymer informatics super-intelligence
}

\author{Akhlak Mahmood}
    \email{akhlak.mahmood@matmerize.com}
    \affiliation{
        Matmerize, Inc., 75 5th Street NW, Atlanta 30308, GA, USA.
    }

\author{Janhavi Nistane}
    \affiliation{
        Matmerize, Inc., 75 5th Street NW, Atlanta 30308, GA, USA.
    }

\author{Huan Tran}
    \affiliation{
        Matmerize, Inc., 75 5th Street NW, Atlanta 30308, GA, USA.
    }
    \affiliation{
        School of Materials Science and Engineering, Georgia Institute of Technology, Atlanta, Georgia 30332, USA.
    }

\author{Chiho Kim}
    \affiliation{
        Matmerize, Inc., 75 5th Street NW, Atlanta 30308, GA, USA.
    }
    \affiliation{
        School of Materials Science and Engineering, Georgia Institute of Technology, Atlanta, Georgia 30332, USA.
    }

\author{Rampi Ramprasad}
    \affiliation{
        Matmerize, Inc., 75 5th Street NW, Atlanta 30308, GA, USA.
    }
    \affiliation{
        School of Materials Science and Engineering, Georgia Institute of Technology, Atlanta, Georgia 30332, USA.
    }

\begin{abstract}

Polymer informatics has matured from isolated property-prediction studies into an integrated discipline that couples data, models, and decision-making across the polymer design cycle. Yet it still falls short of a true intelligent system capable of inverse design on demand, causal reasoning across chemistry, processing, and performance, and closed-loop autonomous experimentation. This article traces a roadmap toward that goal, grounded in experience developing two complementary agentic and informatics platforms. Central to this vision is a modular, agent-directed architecture in which a polymer super-intelligence layer interprets a researcher's design question in natural language and coordinates domain-specialized tools, matched to the available data, for neat polymers, composites and formulations, solvents, and synthesis and processing. The resulting system spans the full chain from molecular design through processing to product-level performance and human perception. Orchestrated together, its generative design, synthesis-feasibility reasoning, and practicality assessment already form the decision-making core of a self-driving polymer laboratory, leaving autonomous, closed-loop experimentation as the principal step that remains. We survey emerging capabilities along this roadmap, including automated extraction of property data from the literature, chemistry-aware representation, property prediction for membranes and sustainable plastics, solubility and green-solvent recommendation, and computer-guided retrosynthetic planning, exposing the remaining gaps and the research and infrastructure investments needed to move from today's orchestrated tool ecosystem toward a genuinely super-intelligent polymer design partner.

\end{abstract}


\maketitle
\doublespacing
\raggedbottom

\section*{Introduction}

Few classes of materials are as versatile as polymers, and few are as hard to design. The same backbone chemistry can be dialed from rigid to elastic, insulating to conducting, or opaque to optically clear by tuning the repeat unit, the composition of copolymers, the processing history, and the fillers and additives that turn a neat resin into a composite or formulation.\cite{lihuaReview,naturalPlasticAL} Every such knob multiplies an already astronomical chemical and processing space, and structure--property relationships are strongly non-linear. Any viable material must also clear competing bars of performance, cost, safety, manufacturability, and, increasingly, sustainability.\cite{phaBioplasticDesign,tropic} Polymer informatics emerged over the past two decades as a data-driven response, on the premise that machine-learning models trained on curated data can learn these relationships and predict the behavior of new candidates without immediate recourse to synthesis and characterization.\cite{huanReview,bayesianMaterialsDesign,stimuliResponsivePolymerAI,mlPolymerResearch} The field has since matured from isolated property-prediction studies into an integrated discipline: models such as polyBERT read the repeat unit directly from a SMILES string and turn it into a \emph{fingerprint}, the numerical summary of structure that a predictive model actually consumes,\cite{polymerGenome,polyBERT} single models trained on dozens of properties at once forecast all of them from structure alone, borrowing strength across properties to stay accurate even when data are scarce,\cite{chrisMultitask,polyOmics} and \emph{inverse design} runs these models in reverse to identify chemistries meeting a target specification, with virtual forward synthesis (e.g. RxnChainer) ensuring the proposed candidates can actually be made.\cite{rxnchainer} Together, prediction and inverse design close a loop that has compressed discovery cycles from decades to months.\cite{aiDielectrics,amActiveLearning}

This closed loop now specializes into recognizable workflows organized by material class, each with a distinct decision logic. For \emph{neat polymers}, it couples literature mining, property prediction, and generative design to propose and rank new repeat-unit chemistries.\cite{polyMetriX,mlPolymerMolecularDesign,petReplacement} For \emph{composites and formulations}, mixture-aware models predict how matrix chemistry, filler loading, and additive composition combine to set properties such as flammability and conductivity.\cite{compositeInformatics,flameRetardant} For \emph{solvents}, the focus shifts to predicting polymer solubility and identifying greener replacements for hazardous solvents.\cite{greenSolvents,llmSolubility} For \emph{synthesis and processing}, retrosynthetic models work backward from a target structure to suggest feasible monomer combinations and polymerization routes.\cite{polyRetro}
Cutting across these domains, large language models (LLMs) have emerged as a versatile layer: they read millions of articles and extract structured property data otherwise locked in text,\cite{llmExtraction} reconcile the inconsistent nomenclature that plagues polymer databases, and, in domain-adapted forms such as polyT5 and polyBART, fold prediction and generation into a single text interface.\cite{polyT5,pointDatabase,benchmarkLLM}

Yet the capabilities remain difficult to use for any researcher who is not simultaneously a polymer chemist and an ML practitioner. A chemist who wants to go from a target property profile to a ranked list of synthesizable candidates today must locate and run several separate tools developed in isolation, manually reformat outputs between steps, and judge which model to trust, since no shared problem-description language or data-curation standard exists. Individual models are also narrow: they typically capture structure and intrinsic properties while overlooking processing, formulation effects, and the product-level performance and end-use perception that ultimately matter.\cite{huanReview} No single system yet combines candidate recommendation on demand, causal reasoning across chemistry, processing, and performance, and closed-loop autonomous experimentation, although these capabilities are beginning to converge.

\begin{figure*}[htbp]
    \centering
    \includegraphics[width=\textwidth]{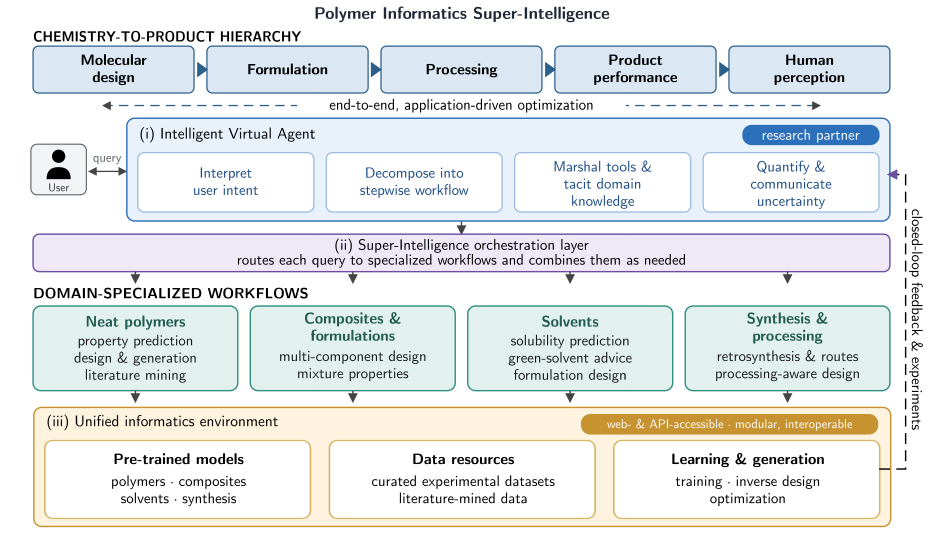}
    \caption{Architecture of the envisioned polymer informatics super-intelligence. (i) A conversational agent interprets the user's natural-language intent and marshals the relevant tools and domain knowledge; (ii) domain-specialized workflows, neat polymers, composites and formulations, solvents, and synthesis and processing, that the orchestrating layer routes queries to; and (iii) a unified, API-accessible informatics environment exposing pre-trained models, data, and algorithms as interoperable building blocks. Closed-loop feedback from users and experiments drives iterative improvement. Early instances of the three numbered layers are already in use while the closed-loop, cross-scale integration that unifies them into a single super-intelligence is the envisioned addition.}
    \label{fig:architecture}
\end{figure*}

Convergence of these capabilities points to the next stage of polymer informatics, what we term a \emph{polymer informatics super-intelligence}: a system spanning the full chemistry-to-product hierarchy, from molecular design through formulation and processing to product-level performance and human perception, capable of end-to-end optimization of complex, application-driven problems no single human or model spans today (see \textbf{Figure~\ref{fig:architecture}}). Such a system would behave less like a calculator and more like a knowledgeable research partner. It would interpret a natural-language design question, decompose it into stepwise tasks, invoke the appropriate tools, quantify its uncertainty, and improve through feedback from users and experiments. Early instances of three principal components are already in use across academic, open-source, and commercial efforts:\cite{polymerAgent,polyJarvis,multiAgentPolymer} (i) conversational agents that understand natural-language queries and select the right tool chain for each task; (ii) domain-specialized workflows that a coordinating agent assembles and sequences, with built-in feedback loops for iterative refinement; and (iii) unified, web- and API-accessible informatics environments that expose pre-trained models, curated data, and design algorithms as interoperable building blocks a chemist can invoke without writing ML code.
Today's most capable systems are \emph{orchestrators}. An LLM agent interprets a design question, selects the right tool for the material class and data regime, and sequences validated tools into an answer. A polymer super-intelligence is this orchestrator extended with the capabilities it still lacks: cross-scale causal reasoning, continual learning from each result, and closed-loop autonomous experimentation, capabilities that set the super-intelligence of tomorrow apart from the orchestration tools and general-purpose AI assistants available today. Orchestrated together, the design-side capabilities it already provides, generating and screening candidates, judging synthesis feasibility, and weighing practical constraints, form the decision-making core of a self-driving polymer laboratory; closed-loop autonomous experimentation is the step that would complete it.

This article traces the progress towards the polymer informatics super-intelligence, grounded in our experience building two complementary platforms that already realize parts of it: \textit{ASKPOLY}, a conversational agent, and \textit{PolymRize}\textsuperscript{TM}, the informatics environment behind it. Both serve as illustrative case studies rather than the subject of the review, chosen because we know their internals in detail; comparable academic, open-source, and commercial systems are cited throughout. The roadmap is intended for a community-wide discourse, since no single group can supply the data, models, and autonomous laboratories the remaining steps demand.

\section*{An intelligent conversational agent}

\begin{figure*}[htbp]
    \centering
    \includegraphics[width=\textwidth]{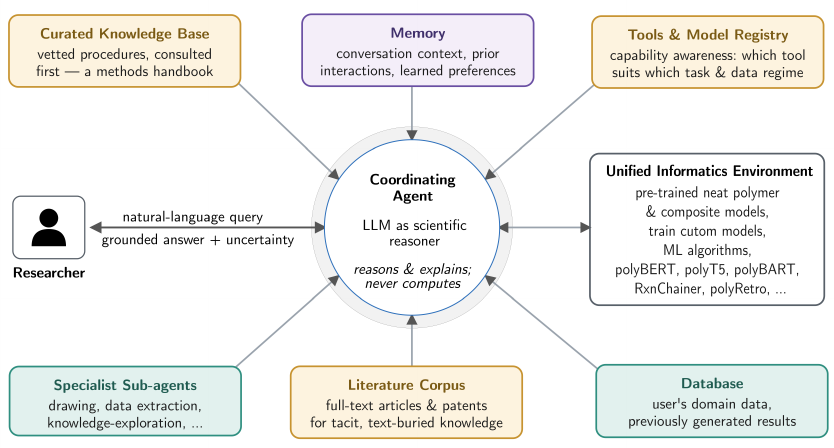}
    \caption{Anatomy of the intelligent conversational agent. At its core, a coordinating agent uses a large language model as a scientific reasoner that interprets intent, plans a solution, and explains the outcome, but never computes numerical results itself. It exchanges natural-language queries and grounded, uncertainty-aware answers with the researcher, and draws on a set of components: a \emph{memory} of conversation context and prior interactions; a \emph{curated knowledge base} of vetted procedures that is consulted first; a \emph{tool and model registry} encoding which tool suits which task and data regime; a \emph{literature corpus} mined for tacit, text-buried knowledge; \emph{specialist sub-agents} (e.g., structure-drawing and knowledge-exploration), each scoped to its own tools; a \emph{database} of the researcher's own domain data and previously generated results; and \emph{validated compute tools} for property prediction with uncertainty, structure rendering, design-space search, and reference lookup---all backed by the unified informatics environment.}
    \label{fig:agent}
\end{figure*}

Consider the everyday experience of a polymer chemist who wants to find a membrane material with high proton conductivity: today that researcher must locate a property database, convert polymer names into a compatible representation, run a predictive model, interpret the output, and manually evaluate candidates.\cite{chemProps} A polymer informatics conversational agent collapses this fragmented workflow into a single interaction. The researcher types the question in plain language; the agent works out which tools are needed, runs them in the right order, and returns an answer with the reasoning and evidence behind it. The agent is not a search engine, and not a chatbot that paraphrases training data, but a coordinator that accepts chemistry questions and dispatches them to validated informatics tools. As \textbf{Figure~\ref{fig:agent}} illustrates, the agent draws on several supporting components: a \emph{memory} of the conversation and past interactions; a \emph{curated knowledge base} of vetted domain procedures; a \emph{tool and model registry} recording which computational tool suits which task; a \emph{literature corpus} of mined research articles; a set of \emph{specialist sub-agents} for focused tasks; and a \emph{database} of the researcher's own data and prior results. All of these rest on the unified informatics environment described in a later section; the design choices behind them are best understood through the systems that demonstrate each one.

\subsection*{Reasoning and tool delegation}
The central challenge in building a reliable polymer informatics agent is that a language model, trained on general text, is a fluent writer but an unreliable calculator; ask it to recall the glass-transition temperature of a specific copolymer and it may return a confident but fabricated number, a failure mode known as \emph{hallucination}. The standard remedy, formalized by the ReAct framework as a practical pattern, is to keep the model a reasoner rather than an oracle: it thinks about what needs to be done, calls an external tool, reads the result, and thinks again, repeating until the question is fully answered, delegating every quantitative step to a validated tool.\cite{react} Because the model's reasoning is written out at each step, the derivation is inspectable, and because the model never produces property values from memory, hallucination in the consequential outputs is controlled.

The power of this approach was demonstrated convincingly by two chemical science agents. ChemCrow equipped a language model with eighteen expert-designed chemistry tools and used it to autonomously plan and execute the synthesis of an insect repellent, three organocatalysts, and a novel chromophore, tasks that required chaining multiple tools over many steps without human correction.\cite{chemcrow} Coscientist took the pattern further into laboratory automation, coupling a language model to internet search, code execution, and robotic hardware to design and optimize palladium-catalysed cross-coupling reactions with little human input.\cite{coscientist}

A polymer super-intelligence inherits this structure and applies it to polymer-specific tasks.\cite{toPolyAgent,polyBench,multiAgentPolymer,polymerAgent} The language model interprets the researcher's question, breaks it into steps, and explains the outcome, while validated polymer informatics tools handle the computation---querying a property database, running a predictive model, or drawing a repeat unit. Knowing how to plan, however, is not enough if the agent lacks reliable domain knowledge to plan with.

\subsection*{Grounding in curated domain knowledge}
\begin{figure*}[htbp]
    \centering
    \includegraphics[width=\textwidth]{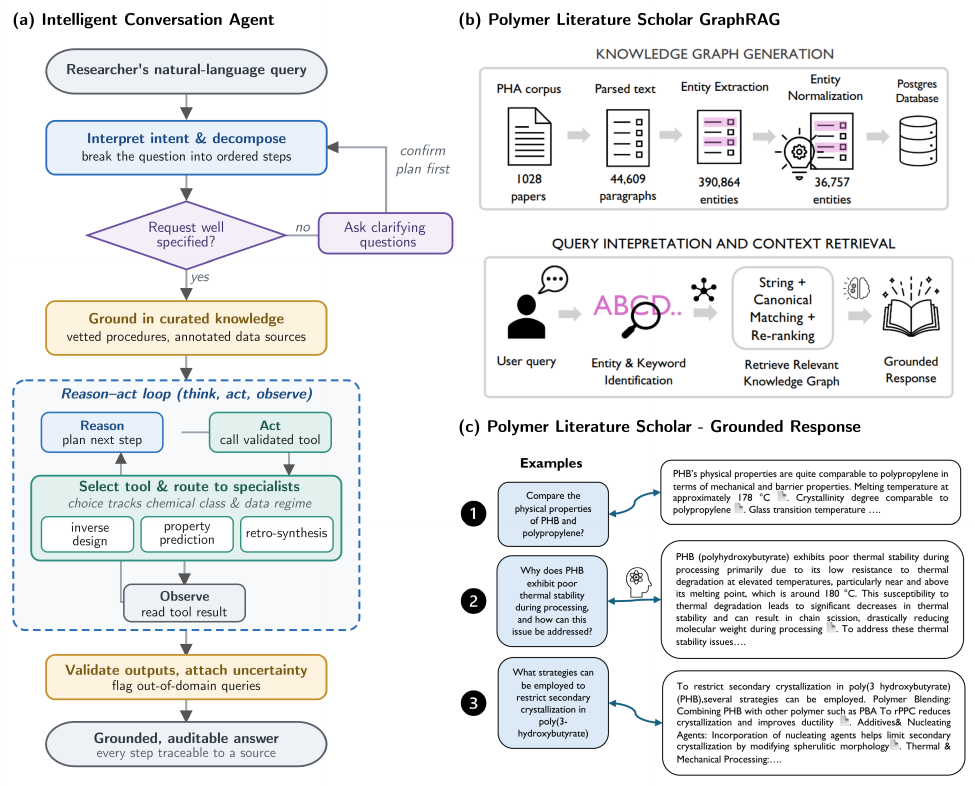}
    \caption{The intelligent conversational agent and its literature-grounding backbone. (a)~Workflow of the agent. A researcher's plain-language query is interpreted and decomposed into ordered steps. If the request is underspecified, the agent asks clarifying questions and confirms the plan before any calculation is run; otherwise it grounds itself in curated domain knowledge and enters a reason--act loop, in the manner of the ReAct pattern, that thinks about the next step, calls a validated tool, and observes the result, repeating until the question is answered. Tool selection routes each sub-task to a purpose-built specialist and is governed by the chemical class and the data regime of the query. Outputs are validated and reported with uncertainty, out-of-domain queries are flagged, and every step remains traceable to its source. (b,c)~The polymer literature scholar built on graph retrieval-augmented generation (GraphRAG). Example grounded responses for queries on polyhydroxybutyrate (PHB), comparing its physical properties to polypropylene, explaining its poor thermal stability during processing, and listing strategies to restrict secondary crystallization, each answer traceable to the underlying literature (taken from Ref.~\cite{polymerRAG}).}
    \label{fig:agent-flow}
\end{figure*}

A general-purpose language model's training carries no polymer design procedures, so an agent must ground its plans in a curated knowledge base of vetted procedures and annotated data sources before invoking any tool.
Several recent systems demonstrate how this grounding works and how much it matters.\cite{honeycomb,llamp} The Polymer Literature Scholar of Gupta and co-workers built a question-answering system over 1{,}028 full-text articles on polyhydroxyalkanoates, split into 44{,}609 paragraph-level passages (\textbf{Figure~\ref{fig:agent-flow}}).\cite{polymerRAG} A knowledge-graph retrieval mode, linking mentions of the same polymer before following relationships across entries, proved considerably more robust than semantic search alone as the corpus grew, retaining a retrieval recall of 0.938 versus 0.717, while both modes held answer accuracy near 0.96--0.97 with each answer traceable to specific source passages. Equally important for research use, the system declines to answer when the evidence is insufficient rather than fabricating a reply, behavior that domain chemists rated as competitive with general-purpose commercial assistants. A challenge specific to the polymer literature is that much of the quantitative record sits not in prose but in tables and figures, often split across several of them within a single article. A retrieval graph built over paragraph-level text alone cannot reach it and extending the graph to link the same polymer across paragraphs, tables, and figures remains an open problem, and the multimodal extraction pipelines that already mine property values from tables and figures are what such a graph would draw on.\cite{llmExtraction,polyie} In a polymer super-intelligence, grounding extends beyond the published literature to the researcher's own accumulated experimental data, so that answers stay informed by the specific materials and conditions at hand and improve as the record grows without retraining the model.

\subsection*{Tool selection across chemical classes and data regimes}
Knowing how to plan and what to trust is only part of being a useful research partner; the rest is knowing which computational tool fits the problem at hand and being honest about the limits of the one selected.
Such an agent therefore maintains a registry of available tools and models, annotated with what each is reliable for and where it is known to break down. Faced with an ambiguous or underspecified request, the agent does not guess: it asks clarifying questions and describes the available approaches so the researcher can confirm the plan before any calculation is run.
Tool selection in polymer informatics turns on two considerations. The first is \emph{chemical class}: neat polymers, multi-component formulations, and small-molecule solvents each require different representations, datasets, and models, so the agent must first establish which family a query belongs to before choosing an approach. The second is the \emph{data regime}: how much reliable experimental data exist for the target property. Physics-based or physics-informed models are more trustworthy when data are scarce; data-driven and generative methods come into their own when data are plentiful; and literature-extraction tools are needed when the relevant knowledge is buried in published prose, tables, and figures rather than in any structured database.\cite{benchmarkLLM}
Raw model size is not a reliable proxy for which approach will work: on the \textsc{PolyIE} benchmark of 146 hand-annotated polymer articles, fine-tuned domain-specific encoders such as MaterialsBERT (NER micro-F1 81.6) outperformed few-shot GPT-3.5 and GPT-4 at extracting polymer property data, demonstrating that the right tool tracks the task and the data regime rather than model scale alone.\cite{polyie}

\subsection*{Coordination of specialized sub-agents}
Even an agent that plans well, grounds itself in vetted knowledge, and selects tools carefully will reach the limits of any single model when the query is complex enough. A question such as ``suggest a biodegradable polyester with a tensile modulus above 2 GPa and a degradation half-life under one year in seawater'' requires drawing, confirming and normalizing candidate structures, running property predictions for both mechanical and degradation behavior, checking synthesizability, and consulting the literature for experimental precedent; no single model handles all of these tasks reliably. The established solution is to divide the work.

This division-of-labor pattern, already established by Coscientist and identified by Ramos and co-workers as an emerging standard for autonomous scientific systems, extends naturally to the polymer domain.\cite{coscientist,llmAgentReview} A general agent manages the conversation and routes each sub-task to an appropriate specialist, such as a structure-drawing assistant that renders repeat units from a name or SMILES string, or the Polymer Literature Scholar discussed above, invoked whenever a query requires synthesizing evidence from many papers.\cite{polymerRAG} This structure also lets the system grow, since a new capability is added as a new specialist without changing the interface the researcher uses. Three nested roles recur across these systems, defined in Box~1: \emph{agent}, \emph{orchestrator}, and \emph{polymer super-intelligence}. Today's systems realize the first two; the cross-scale causal reasoning that completes the third is discussed in the closing sections.

\begin{figure}[htbp]
    \fbox{\begin{minipage}{0.92\columnwidth}
    \vspace{2pt}
    \textbf{Box~1\,$\vert$\ Three nested roles of polymer intelligence}
    \vspace{4pt}

    \noindent\textbf{Agent.} A single language-model-driven unit that reasons over a task and calls tools to carry it out, whether the generalist that fields the question or a specialist for structure drawing, property prediction, or literature synthesis.

    \vspace{3pt}
    \noindent\textbf{Orchestrator.} The generalist in its coordinating capacity, interpreting the design question, deciding the material class and data regime, routing each sub-task to the tool that fits, and assembling the outputs into one validated, uncertainty-annotated answer (today's reality).

    \vspace{3pt}
    \noindent\textbf{Polymer super-intelligence.} An orchestrator that also reasons across scales, connecting a chemistry or processing change to its effect on the final product, learns continually from each result, and runs closed-loop autonomous experimentation (envisioned system).
    \vspace{2pt}
    \end{minipage}}
\end{figure}

\subsection*{Requirements for trustworthy deployment}
A reasoning loop, a grounded knowledge base, a tool-aware registry, and a coordinating multi-specialist structure are only useful to a bench chemist if the resulting system is trustworthy. One that occasionally fabricates property values or silently misapplies a model is worse than no system, eroding confidence even in correct results.
Ramos and co-workers, reviewing the landscape of scientific language-model agents, identify three requirements for trustworthy deployment: grounding answers in verifiable sources, validating outputs against known constraints, and keeping the researcher in the decision loop.\cite{llmAgentReview} Lei and co-workers reach a similar conclusion independently, naming interpretability and hallucination control as conditions for laboratory adoption.\cite{matsciLLM} The paradigm reviewed here is designed to meet these requirements, though reliability in practice still depends on the individual models, data sources, and tools it integrates. Each requirement maps to a testable criterion: domain answers trace to a knowledge-base source rather than parametric memory; uncertainty estimates are checked against held-out measurements, as the calibrated variance of the ring-opening-polymerization model was,\cite{ropEnthalpy} and every prediction carries an applicability-domain flag, with out-of-range queries returning a refusal to provide concrete answers rather than an extrapolation; extracted values carry both source article and extraction method,\cite{llmExtraction} and a repeated query yields the same tool chain and answer, which shared benchmarks make checkable.\cite{polyBench,benchmarkLLM} The agent confirms its plan before acting, predictions remain hypotheses until experiment confirms them, and expert approval stays mandatory wherever a result commits synthesis effort or bears on safety. These are the properties that separate a dependable research partner from a sophisticated autocomplete.

\section*{A unified polymer informatics environment}

\begin{figure*}[htbp]
    \centering
    \includegraphics[width=\textwidth]{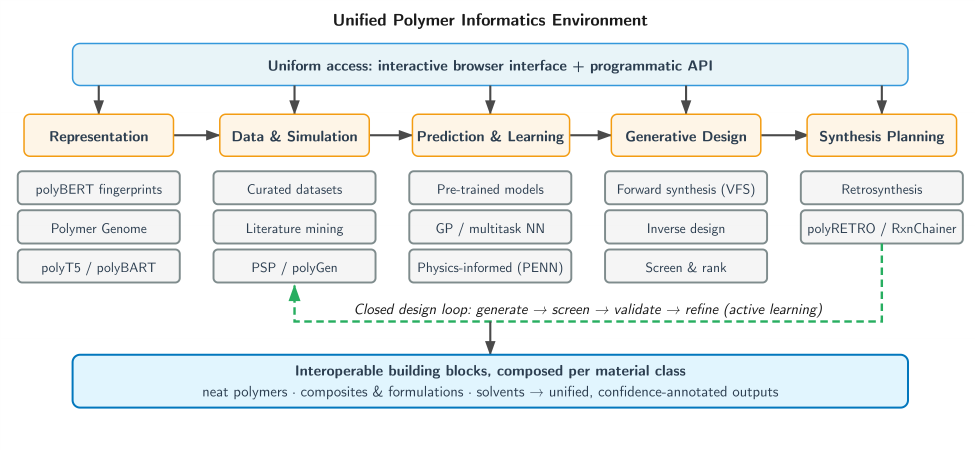}
    \caption{
    The unified polymer informatics environment.
    Every capability of polymer design is exposed as an interoperable building block behind a single uniform interface, reachable both through an interactive browser and through a programmatic API.
    The blocks span the closed design loop: structural representation and fingerprinting (\textsc{polyBERT}, Polymer Genome, \textsc{polyT5}/\textsc{polyBART}); curated data resources and physics-based simulation (literature mining, PSP, polyGen); property prediction and learning (pre-trained models, Gaussian-process and multitask networks, physics-informed models); generative design (virtual forward synthesis, inverse design, screening); and synthesis planning (retrosynthesis via polyRETRO and RxnChainer).
    Because every block shares the same inputs and output format, any two can be chained without reformatting, and the outputs of synthesis and screening feed back as fresh training data in a closed, active-learning loop.
    The same blocks are composed differently for each material class (neat polymers, composites and formulations, and solvents), converging on unified, confidence-annotated outputs.
    }
    \label{fig:routing}
\end{figure*}

The conversational agent is only as capable as the substrate it draws upon. That substrate is the second pillar of the architecture: a unified informatics environment in which the essential tasks of polymer design (encoding structures, curating and mining data, predicting properties, generating candidates, and planning synthesis) are exposed as a coherent, interoperable set of computational tools rather than a patchwork of isolated scripts and packages. Because every tool accepts the same kinds of inputs and returns results in the same format, any two can be chained without manual reformatting, so a researcher can run the pipeline piecewise or an agent can route through it end to end (see \textbf{Figure~\ref{fig:routing}}). A cloud-hosted platform such as \textit{PolymRize}\textsuperscript{TM} is one concrete instantiation, reachable both interactively and programmatically.\cite{polymrize} Crucially, the agent does not route every query through a single fixed pipeline but dispatches it along a material-class-specific path through the shared backend before converging on a unified, confidence-annotated answer. These building blocks differ in maturity, for example, property prediction for the thermal, mechanical, and dielectric behavior of neat polymers is well validated, whereas generative inverse design, retrosynthetic planning, and coverage beyond neat polymers to composites and formulations remain only partially solved. The subsections below discuss the current status and limitations of each block, and the closing section frames the remaining gaps as the field's open problems.

\subsection*{Structural representation}
The environment exposes several interoperable fingerprint families rather than committing to one.\cite{uniPoly,ibmPolymerFM,tgPredictionInformatics} Handcrafted descriptors (Polymer Genome) and \textsc{polyBERT}, which learns its fingerprint from structure data rather than having it specified by hand, both support prediction, with \textsc{polyBERT} matching descriptor accuracy at roughly 200 times the speed, fast enough to screen hundreds of millions of candidates.\cite{polymerGenome,gnnPolymerReview,polyBERT} A fingerprint cannot be run backwards to recover a structure, so \textsc{polyT5} and \textsc{polyBART} pair it with a decoder that writes a structure back out, adding generation to the same representation, and \textsc{BigSMILES} extends the encoding itself to capture an ensemble of chains rather than one idealized repeat unit.\cite{polyT5,polyBART,bigSMILES} A complementary family encodes the repeat unit directly as a molecular graph and learns from it by message passing over atoms and bonds rather than over a fixed fingerprint. The weighted, directed message-passing network of Aldeghi and Coley represents a polymer as a stochastic graph, capturing the monomer stoichiometry and chain connectivity that a single-chain string omits. Additionally, self-supervised pretraining on unlabeled polymer graphs transfers to property prediction where labels are scarce, mirroring the pretraining gains seen for chemical language models; multimodal frameworks such as MMPolymer combine the graph and sequence views to improve prediction across benchmarks.\cite{gnnEnsemble,ssGNNPolymer,gnnPolymerReview,mmpolymer,mipsPolymer} The layer also accepts categorical and continuous descriptors alongside the structural fingerprint, letting composites and formulations be treated on the same footing as neat polymers.

\subsection*{Curated data and high-throughput simulation}
A property model is only as trustworthy as the data beneath it, and for many design-critical quantities the experimental record is too sparse to train on alone.\cite{polyInfoRDF} The environment populates this data layer through literature mining, curated fusion, and simulation, each an agent-invocable step yielding a continuously refreshed resource. Pairing a fine-tuned named-entity model with a general-purpose language model, Gupta \emph{et al.} mined over one million property records for more than 106{,}000 polymers from 681{,}000 articles, roughly twenty times what abstract-only methods yielded.\cite{llmExtraction} Where fidelity rather than coverage is the bottleneck, curated fusion of experimental and computed data fills the gap, as in the multitask Gaussian-process model Toland \emph{et al.} trained for ring-opening-polymerization enthalpy, its calibrated variance flagging which polymers would most benefit from new computation.\cite{ropEnthalpy} Because these sources differ in reliability, the environment records each value's origin and uncertainty rather than discarding the noisier ones. Multi-task learning methods let experimental, simulated, and model-predicted data be combined to extend coverage without a single inconsistent source degrading accuracy, and the model can weight a simulated or predicted value below a direct measurement instead of excluding it.\cite{chrisMultitask,ropEnthalpy,llmExtraction} Where experiments are scarce altogether, physics-based simulation replenishes the record: polyGen and \textsc{RadonPy} automate structure-building and molecular-dynamics calculation, the former roughly two orders of magnitude faster than heuristic builders and the latter computing fifteen properties for more than 1{,}000 amorphous polymers unattended.\cite{predictive3DPrinting,psp,polyGen,radonpy}

\subsection*{Property prediction and learning}
Built on these data, the most immediately useful blocks are the \emph{pre-trained models}. This validated library spans dozens of properties of neat polymers and composites, thermal, mechanical, electronic and dielectric, transport, and thermodynamic.\cite{polymerGenome} Two design commitments make the library usable as a building block: each model returns a calibrated uncertainty alongside its point estimate, reflecting the several sources that make a prediction less certain, sparse training data, variability among the underlying experimental measurements, and a query that falls outside the chemistries the model was trained on, with models trained on several properties at once borrowing strength from the correlations among them to stay accurate where data for any single property are scarce,\cite{chrisMultitask} and every model shares one interface, so predicting one property for one material and predicting many for thousands of candidates are the same operation at different scales. Where no pre-trained model exists, the environment exposes the underlying \emph{learning algorithms}, Gaussian-process and random-forest methods for small data, deep multitask networks for sparse-but-correlated properties,\cite{ropEnthalpy} and physics-informed learning. The physics-enforced network Jain \emph{et al.} built for melt viscosity predicts physically meaningful rheological parameters through a fixed differentiable forward computation, so predictions obey known trends by construction and extrapolate where a physics-agnostic network and a Gaussian process each fail.\cite{deepphysics,pennViscosity} A general-purpose LLM can also serve as a predictor, but how well it works depends on the data regime. Fine-tuned on thermal-property data it matched \textsc{polyBERT}, yet when trained on several properties in sequence it lost accuracy on the ones it had learned earlier. It is therefore offered alongside the fingerprint-based models, not in place of them.\cite{transpolymer,benchmarkLLM} Whichever is chosen, the trained model immediately becomes a new building block indistinguishable in use from the rest.

\begin{figure*}[htbp]
    \centering
    \includegraphics[width=\textwidth]{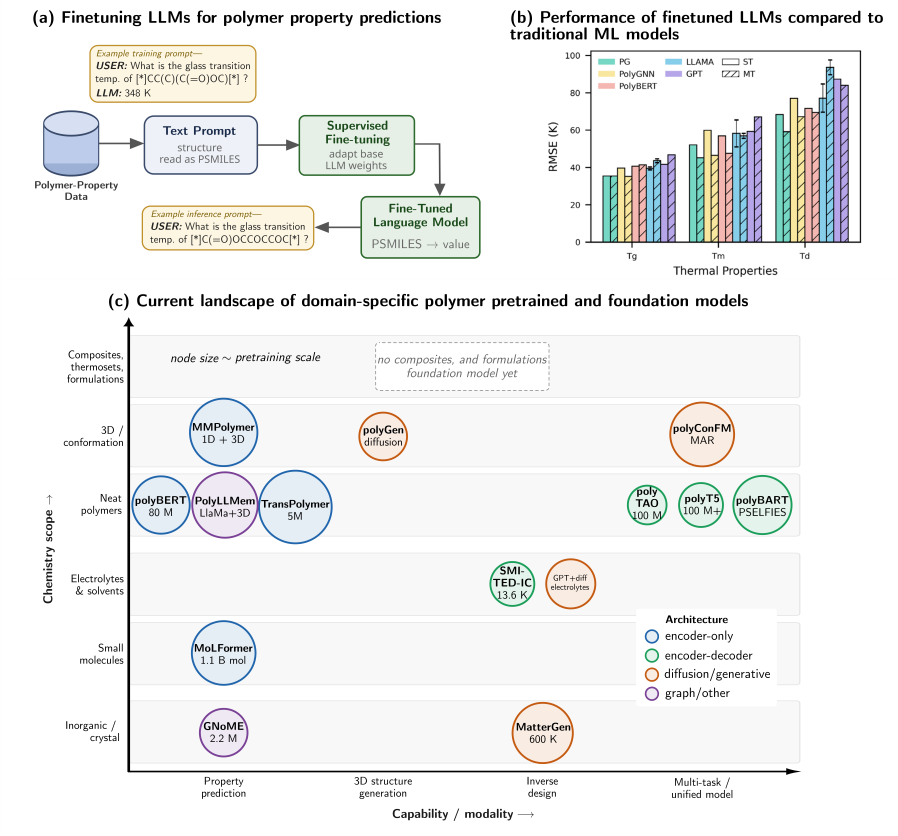}
    \caption{Fine-tuned large language models as polymer property predictors. (a)~Fine-tuning workflow. Polymer-property data are cast as text prompts that encode structure in PSMILES, and supervised fine-tuning adapts the base LLM weights into a fine-tuned model that maps PSMILES to a property value, as illustrated by example training and inference prompts for the glass-transition temperature. (b)~Performance of fine-tuned LLMs compared to traditional ML models taken from Ref.~\cite{benchmarkLLM}. Root-mean-square error (in K) across the thermal properties $T_g$, $T_m$, and $T_d$ for the fine-tuned models (LLAMA, GPT) alongside conventional baselines (PG, PolyGNN, polyBERT), under single-task (ST) and multi-task (MT) settings. (c)~The emerging family of domain-specific pretrained and foundation models for polymer informatics, placed by capability (horizontal: property prediction, three-dimensional structure generation, inverse design, and multitask or unified modeling) and chemistry scope (vertical: from small molecules and inorganic crystals up through neat polymers and conformations to composites, thermosets, and formulations). Node size reflects pretraining scale and color denotes architecture family. The empty upper band marks the open white space.}
    \label{fig:llm-property}
\end{figure*}

The models above are trained one property at a time. A newer generation, \emph{domain-specific foundation models}, is instead trained once on a large collection of unlabeled polymer structures and then adapted to many separate prediction tasks; the library accommodates them as they mature (\textbf{Figure~\ref{fig:llm-property}}). Beyond \textsc{polyBERT}, TransPolymer trained such a model on roughly five million augmented repeat units and improved property prediction across ten benchmarks; the gain came from that unlabeled pretraining step rather than from the network design.\cite{transpolymer} \textsc{polyTAO}, \textsc{polyT5}, and \textsc{polyBART} fold generation into the same network, pretraining on more than one hundred million structures to support inverse design against a specified property target.\cite{polyTAO,polyNC,polyT5,polyBART,selfiesTED,mmpolymer,polyLLMem,polyConFM} Because polymer data carrying measured property labels remain scarce, several efforts transfer instead from molecular foundation models trained at far larger scale, such as MoLFormer, trained on 1.1 billion small molecules; \textsc{polyBART} follows this route through PSELFIES, a polymer-adapted version of the string encoding those models read.\cite{molformer,polyBART}

The scope of these models, however, still stops well short of the chemistry-to-product hierarchy. Coverage is strongest for neat polymers and is beginning to reach electrolytes, where GPT- and diffusion-based generators designed amorphous polymer electrolytes whose best generated repeat units exceeded every ionic conductivity in the training set.\cite{gptDiffusionElectrolyte} Adaptation can also reach the formulation level: SMI-TED-IC fine-tuned the SMI-TED chemical foundation model on 13,666 experimental ionic-conductivity measurements and, through generative screening of multi-component formulations, raised the conductivity of LiFSI- and LiDFOB-based electrolytes by 82 and 172 percent.\cite{smited,smitedIC}

\subsection*{Generation and synthesis planning}
\begin{figure*}[htbp]
    \centering
    \includegraphics[width=\textwidth]{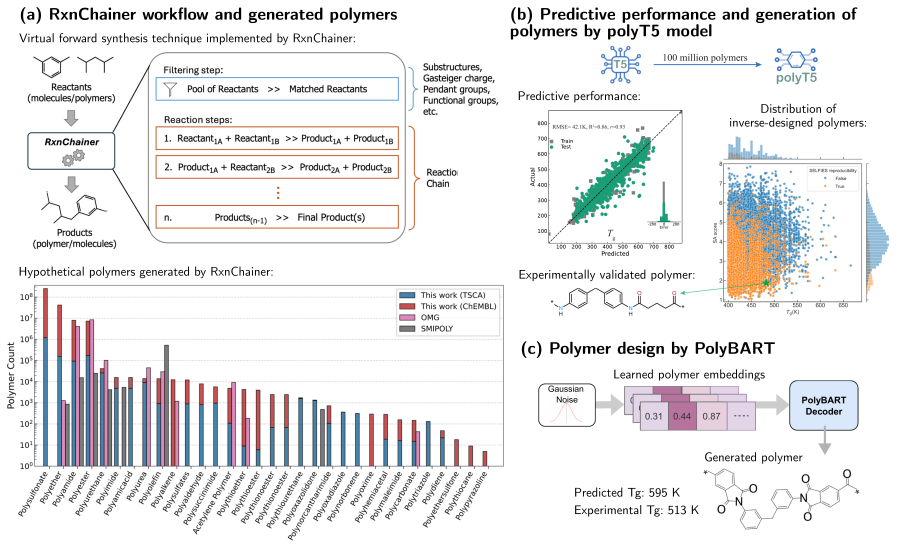}
    \caption{Generative tools for polymer discovery and inverse design. (a)~The RxnChainer workflow for virtual forward synthesis taken from Ref.~\cite{rxnchainer}. Reactant molecules are drawn from a pool and screened in a filtering step using Gasteiger charges, pendant groups, and functional groups, then advanced through successive reaction steps to build a reaction chain that yields final polymer products. The hypothetical polymers generated by RxnChainer (TSCA and ChEMBL) are compared by polymer count against reference sets (DMG, and SMiPoly). (b)~Predictive performance and generation of polymers by the polyT5 model, trained over 100 million polymers, showing a predictive parity plot, the distribution of inverse-designed polymers, and an experimentally validated candidate (from Ref.~\cite{polyT5}). (c)~Polymer design by PolyBART. Gaussian noise is mapped through learned polymer embeddings and a PolyBART decoder to generate a candidate polymer, here with a predicted $T_g$ of 595~K against an experimental value of 513~K (from Ref.~\cite{polyBART}).}
    \label{fig:generative-design}
\end{figure*}

Prediction characterizes chemistries that already exist; discovery requires proposing new ones that can actually be made, so the environment treats synthesizability as an explicit capability. In the forward direction, \emph{virtual forward synthesis} reacts libraries of commercial monomers through curated polymerization templates so every enumerated candidate is makeable by construction. RxnChainer applied 44 reaction chains across 32 polymer classes to filtered monomer inventories, yielding over 289 million hypothetical yet potentially makeable homopolymers, two to three orders of magnitude more than prior enumerations, and a property-driven screen independently rediscovered recently synthesized high-temperature dielectrics.\cite{rxnchainer} \emph{Generative inverse design} instead searches the makeable space toward a target: property-conditioned generators such as \textsc{polyT5} and \textsc{polyBART} produce millions of valid, novel candidates whose predicted properties track the conditioning targets, with experimentally validated dielectric and high-$T_g$ examples confirming the closed loop (\textbf{Figure~\ref{fig:generative-design}}).\cite{polyT5,polyBART} Once a target structure is at hand, \emph{retrosynthetic planning} maps it back to monomers, either by running RxnChainer's templates in reverse or, where symbolic templates are sparse, through a learned route such as polyRETRO.\cite{rxnchainer,polyRetro} Generation, prediction, and synthesis planning thus chain into a define--enumerate--screen--validate loop without leaving the environment.

\subsection*{Composites, formulations, and solvents}
The same building blocks extend natively to composites, formulations, and solvents, which the environment supports on the same footing as neat polymers. A composite's performance depends on the interplay of matrix, fillers, additives, and processing, too intricate for empirical rules such as the rule of mixtures, so the representation layer encodes matrix identity alongside numerical filler loadings and test parameters. Tran \emph{et al.} curated a database of more than 5{,}000 commercial composites and trained physics-informed multitask models that reached a training $R^2$ above 0.9 on twelve of fifteen properties, candidly attributing the exception, breakdown electric strength, to multi-scale physics and missing composition data.\cite{compositeInformatics} Solvents receive the same treatment. Solubility prediction spans the data-regime spectrum, from an LLM classifying solubility directly from a polymer's and solvent's names to tree-based models trained on tens of thousands of controlled turbidity measurements.\cite{llmSolubility,monaSolubility} A structure-based Gaussian-process model assembled a database of over 10{,}000 solvents with predicted green scores and, filtered through a Hansen-solubility criterion, recommends greener substitutes, reproducing known substitutions while proposing novel ones and keeping its uncertainty calibrated as data accumulate, unlike a fine-tuned or in-context GPT-3.5 baseline.\cite{greenSolvents,benchmarkLLM} Kunieda \emph{et al.} likewise fused molecular-dynamics descriptors with interpretable classifiers to predict solvent-resistance failure at ROC-AUC 0.85--0.91 while recovering the governing physics.\cite{screenChemicalResistance}

\subsection*{Uniform access and interoperability}

What makes these blocks a \emph{system} rather than a toolbox is uniform access: browser and programmatic endpoint alike reduce to the same submit-a-task, retrieve-the-result operation, which is what lets an agent act as an autonomous client.\cite{polymrize} Individually, each block answers a focused question; their value as a system lies in how they chain into closed-loop, domain-specialized design workflows.

\section*{Domain-specialized design workflows}

The capabilities reviewed above become most legible when assembled into application-driven workflows and tested against real design targets. Each material class imposes its own defining constraint and therefore calls for a distinct workflow assembled from the shared building blocks (Table~\ref{tab:workflows}).

\begin{table*}[t]
\caption{Representative domain-specialized workflows. Each material class imposes a defining constraint that selects a workflow pattern from the shared building blocks; the orchestrator routes a query to the pattern that fits.}
\label{tab:workflows}
\renewcommand{\arraystretch}{1.4}
\begin{tabular*}{\textwidth}{@{\extracolsep{\fill}}p{0.26\textwidth}p{0.30\textwidth}p{0.34\textwidth}@{}}
\toprule
\raggedright \textbf{Domain} & \raggedright \textbf{Defining constraint} & \raggedright \textbf{Workflow pattern} \tabularnewline
\midrule
\raggedright Neat polymers & \raggedright Closed-loop recycling window plus full mechanical profile & \raggedright Virtual forward synthesis $\rightarrow$ multi-property scoring $\rightarrow$ multi-criterion screen\cite{recyclableROP,recyclablePackaging} \tabularnewline
\raggedright Additive manufacturing & \raggedright Few data, multi-objective composition target & \raggedright Active learning over iterative experimental rounds\cite{amActiveLearning} \tabularnewline
\raggedright Synthesis \& processing & \raggedright Approved-monomer inventories, hazard-free routes & \raggedright Reaction-templating $\rightarrow$ confidence-filtered screen\cite{petReplacement,greenH2} \tabularnewline
\raggedright Separations \& barrier materials & \raggedright Sparse data, long-horizon extrapolation & \raggedright Physics-informed models embedding decay/Arrhenius laws\cite{aemDegradation,solventSeparation} \tabularnewline
\raggedright Composites \& formulations & \raggedright Performance set by recipe, not chemistry alone & \raggedright Models reading the full recipe, including process and architecture, as text\cite{solarCells} \tabularnewline
\bottomrule
\end{tabular*}
\end{table*}

The most demanding neat-polymer targets show the pattern in full. Kern and co-workers sought chemically recyclable ring-opening-polymerization (ROP) polymers that also match polystyrene in mechanical durability.\cite{recyclableROP} Virtual forward synthesis produced 7.3 million synthetically accessible structures; multitask predictive models then scored every candidate against six simultaneous criteria, including a polymerization-enthalpy window of $-10$ to $-20$~kJ/mol that permits closed-loop chemical recycling alongside thermal and mechanical thresholds. A scoring scheme that penalizes failure on any single criterion narrowed the pool to 817 candidates, a hit rate below 0.5\%, of which three were validated in the laboratory, among them a polyamide reaching 2.28~GPa Young's modulus with 93--98\% monomer recovery on recycling. Where data begin to accumulate, active learning closes the loop, retraining the model after each measurement round to choose what to make next: Jain and co-workers used this strategy over six rounds to tune acrylate resins for additive manufacturing, converging within 0.7\% of a target Young's modulus and 3D-printing a single multimaterial part whose rigid ``bones'' ($E \approx 1100$~MPa) and compliant ``skin'' ($E \approx 3$~MPa) were derived from the same three monomers.\cite{amActiveLearning}

Across the other classes the building blocks are the same but their assembly differs with the constraint. When candidates must come from approved monomer inventories or hazard-free routes, reaction-templating engines respect that constraint from the outset, returning shortlists in which every entry already carries a defined synthesis route. Kim and co-workers encoded PET's esterification chemistry as a reaction pattern, applied it to 28{,}289 regulation-listed monomers to enumerate 12{,}100 synthetically accessible copolymers, and after a confidence filter retained 1{,}108 high-confidence structures; the search retrospectively recovered all three established commercial PET alternatives (PETG, Tritan, Ecozen) and two novel candidates were synthesized and confirmed by NMR and DSC.\cite{petReplacement} Tran and co-workers extended the same idea to fluorine-free proton-exchange membranes, enumerating roughly 66 million candidates and scoring them on eleven property targets to yield 1{,}738 finalists validated against in-house Nafion and Pemion measurements.\cite{greenH2} When data are sparse and the target is a long-horizon property such as membrane lifetime, the critical resource is not more data but better use of known physics. Schertzer and co-workers embedded a mechanistic decay law into a neural network that, trained on only 200 hours of anion-exchange-membrane conductivity data, forecast degradation to 10{,}000 hours at $R^2 = 0.987$, where conventional regression predicted the physically impossible result of conductivity rising over time; the same physics-embedded approach, applied to Arrhenius temperature dependence across 3{,}044 polymer--solvent systems, catches where a predicted single-component permselectivity sits orders of magnitude above true mixed-solvent values.\cite{aemDegradation,solventSeparation} Where measurements are plentiful, the same domain instead screens directly for target properties and extracts the governing chemical motifs.\cite{gasSeparationML,interpretableMembraneML} The data regime, not the application, selects the pattern. When performance is set by a full recipe rather than the repeat unit alone, the workflow must absorb processing variables as first-class inputs: Sahu and co-workers adapted a compact language model to predict organic-solar-cell metrics from a natural-language recipe encoding donor/acceptor structures, blend ratio, solvent, thickness, annealing, and architecture, reaching a power-conversion-efficiency $R^2$ of 0.83 on 422 devices and, by screening 1.4 million process-and-architecture combinations, surfacing 2\% DIO additive and ${\sim}125\,^\circ$C annealing as the dominant levers.\cite{solarCells}

\begin{figure*}[htbp]
    \centering
    \includegraphics[width=\textwidth]{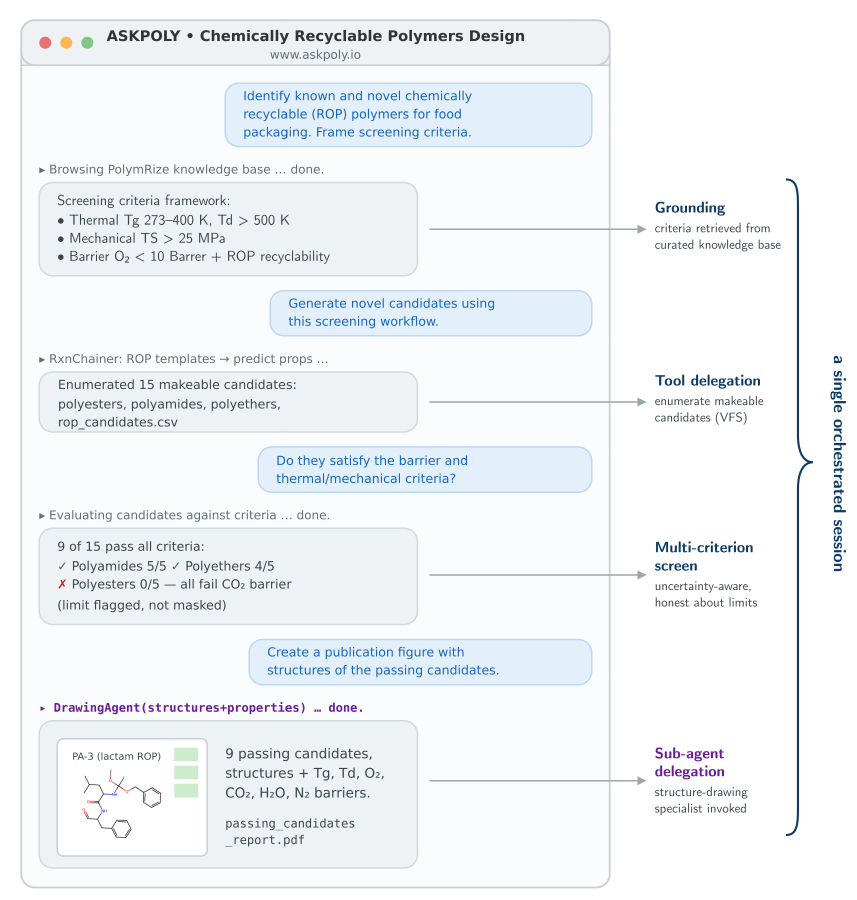}
    \caption{
    A single orchestrated \textit{ASKPOLY} session, reproducing conversationally the recyclable ring-opening-polymerization (ROP) food-packaging workflow of Table~\ref{tab:workflows}.\cite{recyclableROP}
    Through four natural-language turns, the agent (a) \emph{grounds} the problem by retrieving thermal, mechanical, barrier, and recyclability screening criteria from the curated knowledge base;
    (b) \emph{delegates} candidate generation to reaction-templated virtual forward synthesis, enumerating fifteen makeable polyester, polyamide, and polyether candidates;
    (c) applies a \emph{multi-criterion screen}, retaining nine and reporting that all five polyamides pass while every polyester fails the CO\textsubscript{2} barrier limit rather than masking the shortfall; and
    (d) routes structure rendering to a dedicated drawing \emph{sub-agent}.
    The same exchange exercises grounding, tool delegation, multi-specialist coordination, and honest reporting of limits as one continuous interaction, illustrating the orchestrator in use.
    Transcript abridged for clarity.
    }
    \label{fig:askpoly-session}
\end{figure*}

These workflows are instances of one routing discipline, and an early instance of such an agentic framework is already in use. Posed in plain language with the recyclable-ROP food-packaging target of Table~\ref{tab:workflows},\cite{recyclableROP} \textit{ASKPOLY} carries the design loop across four conversational turns (\textbf{Figure~\ref{fig:askpoly-session}}). It grounds the problem by retrieving thermal, mechanical, barrier, and recyclability criteria, enumerates fifteen makeable polyester, polyamide, and polyether candidates through reaction-templated forward synthesis, and predicts their properties. Screening against the assembled criteria, it retains nine candidates: all five polyamides pass, while every polyester fails the CO\textsubscript{2} barrier limit, a shortfall it flags rather than masks. Rendering the surviving structures is then delegated to a dedicated drawing sub-agent, so the single exchange draws on grounding, tool delegation, and multi-specialist coordination.

This routing pattern shows today's orchestrator reasons well about which tool to call, not about the physics those tools encode. It can rank candidates, but it cannot yet trace how a chemistry or processing change propagates to final performance, the capability a polymer super-intelligence must add.

\section*{Outlook and critical next steps}

Carrying today's orchestrated tool ecosystem toward a genuinely autonomous design partner depends less on any single algorithmic advance than on closing a set of concrete gaps. The most critical range from the immediately actionable to the longer-horizon investments the field must make collectively.

\subsection*{Closed-loop autonomous experimentation}
The workflows reviewed in this article still require two manual steps: initiating the next experimental round and returning its results to the model. Closing this loop is the defining transition from orchestrated informatics to a self-driving polymer laboratory, one that would propose a candidate, dispatch synthesis instructions to robotic hardware, receive characterization results, retrain its models, and issue updated recommendations without human handoffs. Early instances already exist, but each is confined to a single reaction class or formulation problem, none yet spanning the full chain from molecular structure through processing to product-level performance.\cite{coscientist} For polymer informatics, where a single cycle spanning synthesis, processing, and characterization may take days, automating these steps would compress the time from candidate shortlist to experimentally validated result from weeks to hours.

\subsection*{Continual learning and adaptive knowledge accumulation}
Closing the experimental loop produces a steady stream of measurements, but today's agents do not learn from it continuously. Retraining is a discrete event a researcher must trigger. A more capable polymer agent would instead update its models as measurements arrive, carry forward validated constraints from earlier sessions, and sharpen its tool-selection judgments over time. This requires uncertainty estimates that remain calibrated as each campaign moves into new chemistry, traceability back to each measurement's origin to know which to trust, and fine-tuning that does not erase performance on previously learned properties. Reaching the non-specialist researchers this capability is meant to serve also depends on the interface: a purely conversational agent shifts the burden of stating and interpreting intent onto exactly the users least equipped to catch a misinterpretation. Hybrid interfaces, pairing natural language with structured controls for routine steps, make the capability usable in practice, provided the agent stays honest about the boundary of its competence.

\subsection*{Toward a general-purpose polymer foundation model}
Cross-scale causal reasoning, the capability that finally separates an orchestrator from a super-intelligence, can either be composed by orchestration, chaining specialized models, simulations, and experimental data from deliberately varied process conditions, which is the route the preceding sections describe, or internalized in a single model. The second defines the most ambitious architectural target for the field: a polymer foundation model, pretrained on a broad, diverse body of polymer data and then adapted to many tasks without building a separate model for each.\cite{stanfordFM} Current models are task-specific. A property predictor for glass-transition temperature knows nothing of synthesis routes, and a generative model trained for dielectrics does not transfer to biodegradable packaging. A foundation model, trained jointly on repeat-unit structures, measured properties, simulation outputs, synthesis routes, processing conditions, and formulation compositions, would internalize the grammar of polymer structure and behavior the way a language model internalizes the grammar of text. One base could be pointed forward to predict a property, run in reverse to generate candidates meeting a specification, or queried for a synthesis route, all without retraining, letting knowledge learned from data-rich properties transfer to data-poor ones. The analogy to language holds only so far; language models learn from trillions of tokens, while the largest curated polymer property collections hold about a million records, each carrying a measurement uncertainty that text does not.\cite{llmExtraction} Scale alone will therefore not suffice; the practical target is a hybrid in which polymer chemistry and physics-based constraints supply the structure that data cannot. A foundation model is also the natural setting in which to resolve a deeper limitation of every predictor reviewed here. Each learns only correlations between inputs and outputs, so it can rank candidates but cannot say how the product responds to a change in a processing parameter. Trained jointly on simulation outputs spanning molecular, meso-scale, and macroscopic behavior, with each manufacturing step represented as a deliberate change rather than an incidental correlation, such a model could distinguish a chemistry change from a process change. Physics-enforced architectures, which encode governing equations as fixed computational layers, are accordingly a core ingredient, though they have so far been demonstrated one property at a time rather than as a connected cross-scale pipeline.

\begin{table*}[htbp]
\caption{Capabilities of today's orchestrator contrasted with those of the envisioned polymer super-intelligence. Each contrast corresponds to one of the critical next steps.}
\label{tab:capabilities}
\renewcommand{\arraystretch}{1.4}
\begin{tabular*}{\textwidth}{@{\extracolsep{\fill}}p{0.24\textwidth}p{0.32\textwidth}p{0.34\textwidth}@{}}
\toprule
\raggedright \textbf{Capability} & \raggedright \textbf{Today's orchestrator} & \raggedright \textbf{Polymer super-intelligence} \tabularnewline
\midrule
\raggedright Tool selection and routing & \raggedright Established\textsuperscript{a} & \raggedright Established\textsuperscript{a} \tabularnewline
\raggedright Reasoning & \raggedright Sequencing which tool runs, and in what order & \raggedright Cross-scale causal (chemistry/process $\rightarrow$ product) \tabularnewline
\raggedright Learning from results & \raggedright Discrete, researcher-triggered retraining & \raggedright Continual, as measurements arrive \tabularnewline
\raggedright Experimentation & \raggedright Human handoff to synthesize and measure & \raggedright Closed-loop, autonomous synthesis and characterization \tabularnewline
\raggedright Scope of a single model & \raggedright One scale, one property at a time & \raggedright Molecule through processing to product-level performance \tabularnewline
\raggedright Model architecture & \raggedright Task-specific predictors and generators & \raggedright Physics-enforced, chemistry-aware foundation model \tabularnewline
\bottomrule
\end{tabular*}

\vspace{3pt}
{\footnotesize \raggedright \textsuperscript{a}The mechanism of tool selection and routing is available; reliably choosing the best-suited model across chemical classes and data regimes remains challenging.\par}
\end{table*}

\begin{figure}[t!]
    \centering
    \includegraphics[width=3.25in]{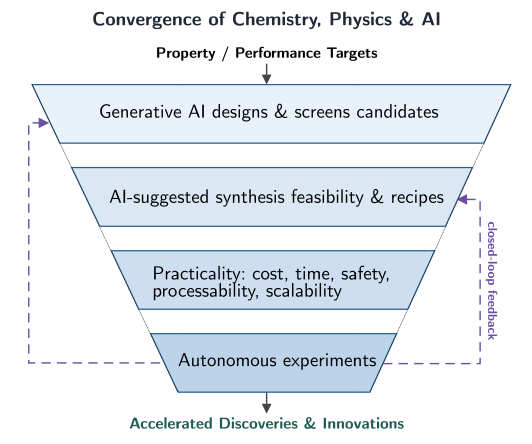}
    \caption{
    The envisioned convergence of chemistry, physics, and artificial intelligence into a closed-loop, autonomous discovery pipeline. The stages correspond to the capabilities reviewed in this article, and the human handoffs still separating them are the critical next steps discussed in this section. Property and performance targets enter at the top. Generative models design and screen candidates, an agent then reasons about synthesis feasibility and recipes and about practical constraints (cost, time, safety, processability, and scalability), and autonomous experiments validate the survivors, yielding accelerated discoveries. Feedback arrows return results from the experiment stage to both generative design and synthesis planning, closing the loop.
    }
    \label{fig:convergence}
\end{figure}

Research toward polymer foundation models is still in its early stages, and the family of pretrained models reviewed earlier (Figure~\ref{fig:llm-property}c) remains confined to neat polymers and conformations, with composites, thermosets, and formulations still unaddressed. The primary barrier is data rather than algorithms. A causal model needs more than a property table, including polymers recorded under a shared nomenclature, so that one material is not split across several names and a chain-length distribution is recorded as a distribution rather than a single idealized chain;\cite{bigSMILES} synthesis routes recorded at the recipe level; processing histories that set morphology; formulation loadings; and an uncertainty estimate and a traceable origin for every measurement. Product-level and perceptual properties such as feel, gloss, and odor, by which a material is finally judged, almost never reach the literature at all. Failed syntheses matter just as much, marking the boundaries of the accessible design space; their absence biases every model trained on the published record toward optimism. OPoly26 released 6.57 million density-functional calculations on polymeric systems under an open license, and CRIPT supplies a shared, cross-institutional schema linking synthesis, processing, characterization, and simulation, though neither yet spans the full chemistry-to-product range.\cite{opoly26,cript,openPolymerChallenge}

These priorities, summarized in Table~\ref{tab:capabilities}, are not independent wishlist items but a connected agenda. Closed-loop experimentation supplies continual learning with the steady stream of fresh measurements it needs, and continual learning is in turn trustworthy only if uncertainty stays calibrated. This is the same property a foundation model needs to tell a genuine cause from a mere correlation and reason across scales. However, all three are starved without a dataset recording recipes, processing, failures, and measurement origin at a scale no single group can assemble alone, making shared, cross-institutional data infrastructure the common prerequisite beneath them. The hardest remaining problems, then, are no longer purely about predictive accuracy on a curated benchmark but about autonomy, traceability, causation, and shared data infrastructure. Brought together, these capabilities describe a single closed loop in which chemistry, physics, and artificial intelligence converge on accelerated discovery (\textbf{Figure~\ref{fig:convergence}}). A property or performance target drives generative design and screening, an agent weighs synthesis feasibility against the practical constraints of cost, safety, and processability, and autonomous experiments validate the survivors, each result feeding back to sharpen the next round. The first three of these stages, orchestrated together, already form the decision-making core of a self-driving polymer laboratory; only the fourth stage, experimental validation, has yet to close on its own, still passing through human hands. A polymer super-intelligence will emerge not from one decisive model but when today's orchestrator gains the capacity to reason causally across scales, proposing, testing, learning, and explaining as it carries a design question from natural language to validated material with progressively less human handoff at each turn.

\section*{Declarations}

\subsection*{Funding}
This work was supported by the National Science Foundation through SBIR Phase I Grant
\#2322108 and by the Office of Naval Research through SBIR Phase I Contract
\#N68335-24-C-0121.

\subsection*{Competing interests}
The authors are employees, affiliates or founders of Matmerize, Inc.,
which owns intellectual property rights related to the \textit{ASKPOLY} and \textit{PolymRize}\textsuperscript{TM} platforms discussed in this article,
including proprietary technology and, potentially, further patent rights.

\subsection*{Data availability}
The data underlying the results discussed here are available in the cited primary publications.

\subsection*{Code availability}
All methods reviewed here are described in the cited primary publications,
which state their own code availability terms.
Access information for \textit{ASKPOLY} and \textit{PolymRize}\textsuperscript{TM} is available at \url{https://matmerize.com}.

\subsection*{Author contributions}
Conceptualization: A.M., R.R. Initial draft and figures: A.M. Manuscript review and edit: A.M., J.N., H.T., C.K., R.R.

\def\bibsection{\section*{References}}
%

\newpage
\subsection*{Figure for Table of Contents Only}
\begin{figure*}[thb!]
\begin{flushleft}
 \includegraphics[width=3.25in]{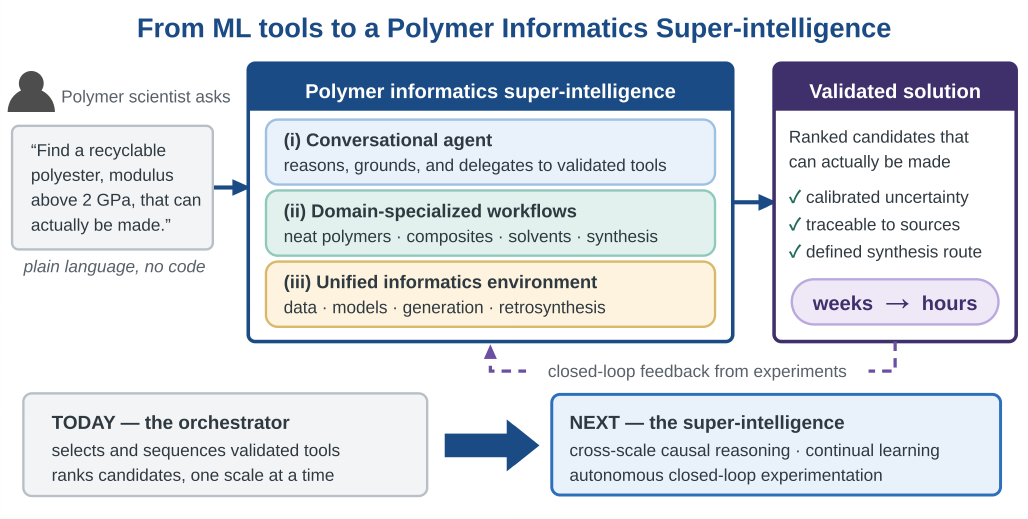}
\end{flushleft}
\end{figure*}

\end{document}